\documentclass[twocolumn]{article}

\usepackage[centertags]{amsmath}
\usepackage{amsfonts}
\usepackage{amssymb}
\usepackage{amsthm}

\date{}
\title{Master Equation for Light emitted by correlated Atoms}
\author{V.N. Gorbachev, A.I. Zhiliba, A.I. Trubilko}
\begin{document}
\maketitle

\begin{abstract}

A master equation for light generated by atoms, which states are
prepared by a pumping mechanism that produces atomic correlations,
is derived in the Fokker - Planck approximation. It has been found
that two-particle correlations only play the role under this
approach. Then the equation is applied for describing micromaser
operations and light noise is discussed. We consider the
correlated atomic states prepared by telecloning protocols.
\end{abstract}

\section{Introduction}

When atoms emit light statistics of radiation depends on the
atomic noises, in particularly, a reason for which is correlation
of atoms. To generate light atomic ensemble can be prepared by the
various pumping mechanisms. As well known, in the case of a
regular pumping atoms may generate light, which statistics of
photons is sub Poisson as for the laser \cite{GolSok},
\cite{Walls} and the micromaser \cite{Golubev} operations. Thus in
the case of micromaser generation sub Poisson statistic of light
can be achieved if atoms, are injected into cavity in such a way
that only $N$ atoms, where $N$ is a small and not random value,
interact with a cavity mode and contributions of atoms are assumed
to be independent. Recently it was found that the quasi-stationary
micromaser operation by N independent atoms has nontrivial
dynamics \cite{MVadeikoRubinT}

The main aim of this paper is to obtain master equation for
radiation, when one cavity mode interacts with atoms which state
is prepared by a pumping that creates correlations between atoms.
Next, as example, we consider how statistics of generated light
depends from atomic correlations. For this we derive a field
master equation for the Glauber-Sudarshan quasiprobability in the
Fokker - Planck approximation to describe interaction of one
cavity mode with $N$ two-level correlated atoms. It is not the
case for using the classical Scully-Lamb approach \cite{LSc} (see
also its application in the micromaser theory \cite{FilLugSc})
because it starts from the single atom task.

States of correlated atoms can be prepared by different ways. In
particularly, any pumping mechanism can be considered as a cloning
process, which output is a set of $N$ copies of an initial atomic
state having an inversion of population of working levels. In
quantum information processing it is already well known the
protocols of obtaining $N$ copies of quantum states from the $M$
originals. It is the transformations $M\to N$, where $M\leq N$.
First of all this protocol can be achieved by a set of unitary
transformations, known as Universal Quantum Cloning Machines
(UQCM) \cite{BuzekHill}, \cite{BuzBraunHill}, \cite{Gisin}. Next,
combining UQCM and teleportation, it has been proposed telecloning
\cite{Vedral}, that allows a producing of spatially separated
copies from originals by the teleportation protocol. Let $M=1$,
then for a pure state of a qubit with the help of UQCM or
telecloning one can obtain the $N$ - particle entangled state of
the GHZ (Greenberger - Horne - Zrilinger) class
\begin{equation}\label{1}
\alpha |0\rangle+\beta|1\rangle \to \alpha|0\rangle^{\otimes
N}+\beta|1\rangle^{\otimes N}
\end{equation}
where $|\alpha|^{2}+|\beta|^{2}=1$, $|b\rangle^{\otimes N}$ is the
tensor product $|b\rangle\otimes\ldots\otimes |b\rangle$, $b=0,1$.
In the same time copies of some quantum states can be created by
teleportation without UQCM, when a multipaticle quantum channel is
permitted \cite{Braunst}. By the multiparticle channel, also named
the multiuser channel, an initial state is distributed with $N$
parties, each of which has an optimal copy of the original in his
hands. It is kind of telecloning. If the original is a mixed
state, such type telecloning allows preparing the $N$ - particle
mixture \cite{Ourprep}
\begin{equation}\label{2}
\lambda_{0}|0\rangle\langle 0|+\lambda_{1}|1\rangle\langle 1|\to
\lambda_{0}(|0\rangle\langle 0|)^{\otimes N}+\lambda_{1}(|1\rangle\langle 1|)^{\otimes N}
\end{equation}
where $\lambda_{0}+\lambda_{1}=1$. Note that in contrast to (\ref{1}), the
obtained $N$-particle state (2) is separable.

According to the no-cloning theorem \cite{Zurek} all copies are
not independent, except the case $\alpha=0$ or $\lambda_{0}=0$,
when the state of $N$ independent particles of the form $
|1\rangle^{\otimes N}$ is obtained. Let $|0\rangle$, and
$|1\rangle$ are the lower and the upper atomic level, then by
telecloning one obtains the correlated states of $N$ atoms. The
collection of these atoms can generate light, if they have
inversion of the population of the working levels, when
$|\beta|^{2}> |\alpha|^{2}$, $\lambda_{1}>\lambda_{0}$. Both
states, given by (\ref{1}) and (\ref{2}), are quite different,
but, with respect to any set of the $K<N$ particles their features
are similar, if $N>2$. The $K$-particle density matrix has the
form $f(12\dots K)=r(|0\rangle\langle 0|)^{\otimes
K}+(1-r)(|1\rangle\langle 1|)^{\otimes K}$, where $r=\lambda_{0},
|\alpha|^{2}$. It means, that any atom has no coherence on the
working transition, its state is mixed, and atomic correlations
are purely classical. Next we also consider the state
\begin{equation}\label{3}
|Z\rangle=\alpha|bb\rangle+\beta|\Psi^{+}\rangle
\end{equation}
where $|\alpha|^{2}+|\beta|^{2}=1$,
$\Psi^{+}=(|01\rangle+|10\rangle)/\sqrt{2}$ is one of the Bell
states, $b=0,1$. This is entangled state, for which  the
one-particle density matrix reads
$f(1)=|\alpha|^{2}|b\rangle\langle b|+|\beta|^{2}\cdot
1/2+(\alpha\beta^{*}|b\rangle\langle
1|\delta_{b0}+|b\rangle\langle 0|\delta _{b1}+h.c.)/\sqrt{2}$.
Indeed, there is the atomic coherence, or polarization, that
results in some new features as, for example, the laser and
micromaser operation without inversion of population on the
working levels \cite{Lmcoh}.

What is a difference between the presented correlated states,
prepared by telecloning, from the states, produced by  any "usual"
pump? As for the simplest model of a pump, one can consider a
collection of $N$ independent atoms, say in the state $f(1\dots
N)=f(1)^{\otimes N}$, $f(1)=\lambda_{0}|0\rangle\langle
0+\lambda_{1}|1\rangle\langle 1|$. Let atoms are illuminated by a
strong coherent field. This interaction can be described by a
unitary transformation, that results in the inversion of
population of the working level. But one finds, that any
correlation between atoms cannot be produced. Then atoms become
active, but independent, its entropy has the form $S=NH_{1}$,
where
$H_{1}=-\lambda_{0}\log\lambda_{0}-\lambda_{1}\log\lambda_{1}$.
The entropy of the pure states (\ref{1}) and  (\ref{3}) is equal
to zero, the entropy of the mixed state (\ref{2}) is $S=H_{1}$.
Because under an unitary transformation entropy does not change,
the correlated states, we consider, can't be prepared by a simple
way, when independent atoms interact with a classical field.

To derive the master equation, we use formalism  presented in
\cite{GZ} for considering the interaction of $N$ correlated atoms
with a quantum electro-magnetic field. In this approach we use the
standard procedure of illuminating of atomic variables assumed to
be fast. It results in the Fokker - Planck equation (FPE) as
example for the Glauber - Sudarshan quasiprobability widely used
for the quantum optics problems. In spite of the states (\ref{1}),
(\ref{2}) and (\ref{3}), we discuss here, describe entangled and
correlated atoms, in fact it is not a real interaction between
atoms and they are treated as a non interacting set of particles.
So that the correlation gives no additional difficulties in
deriving the FPE. One of the main result followed from the used
approximation is that all diffusion coefficients of the obtained
FPE involve the two-atom correlations only. Then for our purpose
it needs consider not less then 2 atoms, interacted with the
cavity modes. In the same time we assume the number of atoms is
limited, to consider the scheme close to micromaser. It is in
accordance with the standard approximation of small fluctuation of
field that we also use. The reason is that, the diffusion
coefficients of the FPE have the contribution proportional to
$N(N-1)$, due from the initial multiatom problem, we consider
here. Under the small fluctuation approximation the diffusion
coefficients must be small, that is the restriction on the number
of atoms.

Using the found FPE, for the case of initial state of atoms (3) we
calculate the Mandel parameter of the light and its spectrum of
noise for the micromaser operation. We find here that atomic
correlations lead to increasing of the the light noise, but in
particular case the noise can be reduced up to zero, if all atoms
are prepared in the upper states $|1\rangle ^{\otimes N}$. Note
that the perfect reduction of the micromaser noise can be achieved
also by the regular pumping as it has been demonstrated in ref.
\cite{Golubev}, when the pump results in the Fock state of the
light. However this generated state has no well defined phase
because of its diffusion. In contrast, if atoms are prepared in
the entangled state, given by (\ref{3}), we found a regime in
which the light is generated without inversion of population of
atomic levels and in particularly, the generated mode has
steady-state phase and noise suppressed up to zero.

\section{Initial equations}

Let consider a collection of $N$ similar two level atoms
interacting with one cavity mode having a frequency equaled to an
atomic transition frequency. This system is described by the
density matrix $F$ that obeys the following equation
\begin{eqnarray}\label{11}
\frac{\partial}{\partial t}F=[\vartheta, F]&&\\
\nonumber
\vartheta=-i\hbar^{-1}V&&\\
\nonumber V=-i\hbar g(S_{10}a-S_{01}a{\dagger})&&
\end{eqnarray}
where  $g=d\hbar^{-1}\sqrt{(\hbar\omega/(2L^{3}\epsilon_{0})}$ is
the coupling constant, $d$ is a dipole moment of a working
transition, $a^{\dagger}$, $a$ are photon operators, $[a;
a^{\dagger}]=1$; atomic operators $S_{xy}$ are defined as
\begin{eqnarray}\label{12}
S_{xy}=\sum_{k=1}^{N}s_{xy}(k)&&\\
\nonumber
s_{xy}(k)=|x\rangle_{k}\langle y|&&\\
\nonumber
x,y=0,1&&
\end{eqnarray}
where $|0\rangle_{k}$ is a lower and $|1\rangle_{k}$ is a upper
state of an atom $k$. The relaxation processes can be taken into
account by adding appropriate terms in (\ref{11}).

To derive a master equation for field based on the formalism
presented in \cite{GZ}, let introduce the representation of the
density matrix $F$ over the coherent states of field
\begin{eqnarray}\label{121}
F=\int d^{2}\alpha~ \Phi(\alpha)|\alpha\rangle\langle\alpha|&&
\end{eqnarray}
where $a|\alpha\rangle=\alpha|\alpha\rangle$. The  matrix $\Phi$
is the operator with respect to the atomic variables but a
c-function with respect to field. When $\Phi$ is averaged over the
atomic variables, one finds the well-known Glauber-Sudarshan
quasiprobability $P(\alpha)=Sp_{A}\Phi$. An equation for $\Phi$
can be obtained from (\ref{11}) taking into account the
correspondence $ [\vartheta,F]\leftrightarrow
[\vartheta_{0},\Phi]+\partial_{\alpha}(D\Phi) $ where field
operators $a$ and $a^{\dagger}$ are presented by complex numbers
$\alpha, \alpha^{*}$ associated with the field amplitudes, and by
the derivatives over these amplitudes as $\partial_{\alpha},
\partial_{\alpha^{*}}$. The Hamiltonian $\vartheta_{0}$ represented the interaction
of the atoms with a field of the given complex amplitude $\alpha$
has the form $\vartheta_{0}=g(S_{01}\alpha^{*}-c.c.)$, the term
involved the derivatives reads $
\partial_{\alpha}(D\Phi)=-g
{\partial}(S_{01}\Phi)/{\partial \alpha}+ h.c. $

Introduce the representation for $\Phi$
\begin{equation}\label{15}
\Phi=P\otimes f+\Pi
\end{equation}
where $f$ is the atomic density matrix for which $Sp_{A}f=1$,
where subscript $A$ denotes spur over the atoms, $\Pi$ is a
correlation matrix, $Sp_{A}\Pi=0$. The field quasiprobability
$P=Sp_{A}\Phi$ obeys the following equation
\begin{equation}\label{16}
\frac{\partial}{\partial t}P=
-g\frac{\partial}{\partial\alpha} \left\lbrack \langle S_{01}\rangle P
+Sp_{A}(S_{01}\Pi) \right\rbrack+h.c.
\end{equation}
where the angle brackets here and later denote the average of the
atomic operators over the density matrix $f$ as $\langle
S_{xy}\rangle=Sp(S_{xy}f)$ $x,y=0,1$. It is worth paying attention
that for $f$ we get here the closed problem in which the
interaction of atoms with field is described by the Hamiltonian
$\vartheta_{0}$, where the field is given by its complex amplitude
only. It looks as the problem of evolution of the atomic ensemble
in a classical field
\begin{equation}\label{17}
\frac{\partial}{\partial t}f=[\vartheta_{0},f]
\end{equation}
Using (\ref{11}) and taking into account (\ref{16}) and
(\ref{17}), one finds the precise equation for the correlation
matrix $\Pi$. However we wish the FPE for the field
quasiprobability $P$. Therefore it needs a solution for  $\Pi$ in
a perturbation theory over derivations to $\alpha, \alpha^{*}$ up
to the first derivatives only, because it gives a term to the
equation for $P$ (\ref{16}) resulting we will have derivates of an
order not higher then two. It results in the following equation
for $\Pi$
\begin{eqnarray}\label{18}
\frac{\partial}{\partial t}\Pi=[\vartheta_{0},\Pi]
-g \left\lbrack \frac{\partial}{\partial \alpha}(S_{01}-\langle S_{01}\rangle)f+h.c \right\rbrack P&&
\end{eqnarray}

To derive a closed equation equation for $P$ usually it is assumed
that an atomic system evolves more faster then field and
introduced a rough time scale. Let $T$ is a typical time of
changing of the field and consider the rough derivative over $T$
as $(P(T+t)-P(t))/T=\partial P/\partial t$. Then integrating
equation (\ref{16}) for $P$ over time from $t$ till $T+t$ and
taking into account the fact, that the field is slowly changed in
the scale time of $T$, we find the desired equation for $P$
\begin{eqnarray}\label{19}
\nonumber
 \frac{\partial}{\partial t}P(t)=&&\\
 \nonumber
=-\frac{g}{T} \int_{t}^{T+t} dt' \frac{\partial}{\partial\alpha} (\langle
S_{01}(t')\rangle P(t) + Sp_{A}(S_{01}\Pi(t'))&&\\
 +h.c.&&
\end{eqnarray}
This is the master equation for field. Its coefficients can be
found from the the problem for $f$ and $\Pi$ where the field is
presented by its quasiprobability $P$ and assumed to be constant
in time. Also it needs to denote the initial matrices $f(0)$ and
$\Pi(0)$ are taken at an initial moment of time, say $t=0$ and it
is used as a way to introduce the pumping mechanism. As for
example, consider atoms placed into cavity, which Q-quality is
modulated periodically, and atoms interact with the high Q cavity
mode during time $T$. Then we can describe a pumping as initial
state of atoms, prepared at time $t$ as $f(t)$, and for an
appropriate choosing of the time intervals we set $\Pi(t)=0$. Also
our scheme can be considered as micromaser, when a beam of atoms,
that states are given by pumping, is injected into cavity, in
which $N$ atoms interacts with one mode during time $T$ and after
leave the cavity. As result, the master equation describes the
field behavior, arisen from summing over the all changing of the
field in time $T$. Note, all contributions are independent.

\section{The atomic averages}

According to (\ref{19}), evolution of the electromagnetic field
depends on the atomic observables of two types. They are based on
the atomic matrix $f$ and the correlation matrix $\Pi$, like
$\langle S_{p}(t)\rangle =Sp_{A}(S_{p}f(t))$ and
$Sp_{A}(S_{p}\Pi(t))$, where $p=0,1,2,3$ or $p=00,01,10,11$, if
the binary notation of $p$ is used. Because atoms are assumed to
be identical, one finds, that $\langle S_{p}(t)\rangle=N\langle
s_{p}(t;1)\rangle $, and the one-particle operator
$f(t;1)=Sp_{2\dots N}f(t;1\dots N)$ is needed. The matrix $f$ can
be obtained by solving  (\ref{17}), that looks as a problem of
atoms in the field of the given complex amplitude $\alpha$, in
other hand it can be treated as a task of interaction of atoms
with a "classical" field. The matrix $f$, as well as $\Pi$, cannot
be factorized, because of the pumping, we consider, prepares the
correlated atoms, in contrast the case of independent particles,
discussed in Ref \cite{GZ}, when both $f$ and $\Pi$ are product of
the one-particle operators. It is important, that there is no any
interaction between atoms, in spite of the atomic correlations are
created by the pumping. For the problem (\ref{17} it results in
operator evolution has the factorized form and one finds
\begin{eqnarray}\label{31}
f(t;1\dots N)=U^{\otimes N}f(t=0;1\dots N)(U^{\dagger})^{\otimes N}&&
\\
 \nonumber
U= \mu+\nu s_{10}(1)-\nu^{*}s_{01}(1)
\end{eqnarray}
where $\mu=\cos(g|\alpha|t)$, $\nu=-(\alpha/|\alpha|)\sin(|\alpha|t)$.

There is another situation for calculating the observables
associated with the correlation matrix $\Pi$. Integrating
(\ref{18}), one finds that all values of the type
$Sp_{A}(S_{p}\Pi(t))$ depend from the atomic variances involving
pair of the atomic operators
\begin{eqnarray}\label{21}
\nonumber
Sp_{A}(S_{p}\Pi(t))=&&
\\
\nonumber
 = -g\int_{0}^{t} dt' \left\lbrack \frac{\partial}{\partial
\alpha}D_{p~01}(t,t')+ \frac{\partial}{\partial \alpha^{*}}D_{10 p}~(t',t)
\right\rbrack P(t)&&\\ &&
\end{eqnarray}
where variances $D_{pq}(t,t')$, $p,q=0,1,2,3$ have the form
\begin{eqnarray}\label{22}
D_{pq}(t,t')=
\langle S_{p}(t) S_{q}(t')\rangle-\langle S_{p}(t)\rangle\langle S_{q}(t')\rangle&&
\end{eqnarray}
where the Heisenberg operators are introduced as
$S_{p}(t)=(U^{\dagger})^{\otimes N}S_{p~}U^{\otimes N}$ and
averaging is taken over the initial state of atoms $f(t=0;1\dots
N)$. Note, that in (\ref{21}) we suggest, that the atomic
operators commute with all derivatives over $\alpha$. That is a
typical approximation for the FPE \cite{GZ}. The variances, given
by (\ref{22}), consist of two parts, where first is the
one-particle contribution proportional to $N$, and second is the
two-particle contribution proportional to $N(N-1)$
\begin{eqnarray}\label{221}
D_{pq}(t,t') =N [\langle s_{p}(t;1)s_{q}(t';1)\rangle &&\\
\nonumber
 -
\langle s_{p}(t;1)\rangle\langle s_{q}(t';1)\rangle ] &&\\
+N(N-1) [ \langle s_{p}(t;1)s_{q}(t';2)\rangle &&
\\
- \langle s_{p}(t;1)\rangle \langle s_{q}(t';1)\rangle ]
 \label{23}&&
\end{eqnarray}
Note, in the classical Lamb-Scully approach one term from
(\ref{221}) $N \langle s_{p}(t;1)s_{q}(t';1)\rangle $ takes place.
For calculating the atomic variances, the Heisenberg picture is
suitable, it results in evolution of the one-atom operator in the
form
\begin{equation}\label{231}
s_{p}(t;1)=\sum_{q}R_{pq}(t)s_{q}(1)
\end{equation}
where $p,q=0,1,2,3$, and the unitary matrix $R_{pq}$ reads
\begin{eqnarray}\label{24}
R_{pq}= \left(
\begin{array}{cccc}
 \mu^{2}    & -\mu\nu  &-\mu\nu^{\ast}   & |\nu|^{2} \\
 \mu\nu     & \mu^{2}  & -\nu^{2}     &-\mu\nu     \\
 \mu\nu^{\ast} &-\nu^{\ast2} & \mu^{2}      &-\mu\nu^{\ast}  \\
|\nu|^{2}   & \mu\nu   & \mu\nu^{\ast}   & \mu^{2}
\end{array}
\right)
\end{eqnarray}
Then all variances take the form
\begin{eqnarray}\label{36}
D_{pq}(t,t')=\sum_{PQ}R_{pP}(t)R_{qQ}(t') D_{PQ}
\end{eqnarray}
where $p,q,P,Q=0,1,2,3$ and $D_{PQ}=D_{PQ}(0,0)$ is an initial
correlation function, given by  (\ref{22}), in which $t=t'=0$. It
can be seen, that the two-particle density matrix $f(t=0;12)$ is
needed to find these variances.  It means, that  the two-atomic
correlation only is required to specify the field behavior under
the Fokker - Planck approximation.

\section{The Fokker-Planck equation and noise}

After introducing the polar coordinates $I=|\alpha|^2$,
$\varphi=\arg \alpha$ the master equation for field (\ref{19})
takes the FPE form
\begin{eqnarray}\label{41}
\frac{\partial}{\partial t}P=\left\lbrack \sum_{u,v=I,\varphi}
\frac{\partial}{\partial u}A_{u}+ \frac{1}{2}\frac{\partial^{2}}{\partial u\partial v}
Q_{uv}\right\rbrack P
\end{eqnarray}
All coefficients $A_{\mu}$ at the first derivatives are denoted
by the one-particle atomic polarization $\langle
s_{01}(t;1)\rangle $
\begin{eqnarray}\label{42}
A_{I}=-N\frac{g\sqrt{I}}{T}\int_{0}^{T}dt\exp(-i\varphi)\langle s_{01}(t;1)\rangle+h.c.&&\\
\nonumber A_{\varphi}=N\frac{g}{2\sqrt{I}T}\int_{0}^{T}
dt~i\exp(-i\varphi)\langle s_{01}(t;1)\rangle+h.c.&&
\end{eqnarray}
At this step the field relaxation has to be taken into account,
because the light must leave the cavity. For the model of the
Lindblad type, it results in $A_{I} \to A_{I}+CI$, where $C\ll
1/T$ is a rate at which photons come out of the cavity. $C$
depends on the output mirror transmitting. The diffusion
coefficients are based on the atomic variances and involve the
two-particle correlations
\begin{equation}\label{43}
 Q_{uv}=\theta_{uv}\frac{g^{2}}{T}\int_{0}^{T} dt\int_{0}^{t} dt'q_{uv}(t,t')
\end{equation}
where  $\theta_{uv}=\theta_{vu}$, $\theta_{I\varphi}=1$,
$\theta_{\varphi\varphi}=1/(4I)$, $\theta_{II}=I$,
\begin{eqnarray}\label{44}
q_{I\varphi}=q_{\varphi I}=
-i\exp(-2i\varphi)D_{11}(t,t')+c.c.&&\\
\nonumber
q_{\varphi\varphi}
=D_{21}(t,t')-\exp(-2i\varphi)D_{11}(t,t')+c.c.
&&\\
\nonumber
q_{II}
=D_{21}(t,t')+\exp(-2i\varphi)D_{11}(t,t")+c.c.&&
\end{eqnarray}

To describe nose of the light we use the Mandel parameter $\xi$,
that indicates how statistics of photons deviates from the Poisson
statistics: $\langle n^{2}\rangle-\langle n\rangle^{2}=\langle
n\rangle(1+\xi)$, where $n=a^{\dagger}a$ is the photon number
operator. In the presented representation $\xi=\langle
\epsilon^{2}\rangle/\langle I\rangle$, where $\epsilon=I-\langle
I\rangle$ is a fluctuation, that is the difference between
"intensity" of the light $I$ and its average value $\langle
I\rangle=\int d^{2}P(\alpha)|\alpha|^{2}$. Assuming, that
fluctuation of intensity is small and does not depend on the phase
fluctuation, then it follows from (\ref{41}) immediately, that the
Mandel parameter is proportional to the diffusion coefficient
$Q_{II}$ as $\xi=Q_{II}(\langle I \rangle \Gamma)^{-1}$, where
$\Gamma=C+(\partial A_{I}/\partial I)_{I=\langle I\rangle}$ is the
decay rate of the intensity fluctuations. The considered
approximation of small fluctuation is the standard one for
analysis the FPE and it needs $\xi\ll\langle I \rangle$.

\section{Noise of micromaser operation}

To apply the obtained FPE consider the micromaser operation, when
the atomic relaxation can be neglected. It is true, if the decay
rates of atomic levels $\gamma$ such that $ 1/T \gg \gamma \gg C$.
Assume  fluctuations of intensity are small in a sense, that the
deviation $\epsilon =I-\langle I\rangle\ll\langle I\rangle$
slightly change near a steady state solution $\langle I\rangle $.

First consider the pump, produced classically correlated atoms in the state,
given by (\ref{1}), if $N>2$ and (\ref{2}). Then from (\ref{41}) it follows,
that $\langle I\rangle$ is the steady state of the semiclassical micromaser
equation of the form
\begin{equation}\label{51}
 \frac{\partial}{\partial t}\langle I\rangle
 =-C\langle I\rangle+\frac{N}{T}(\lambda_{1}-\lambda_{0})
 \sin^{2}B
\end{equation}
where $B=g\sqrt{\langle I\rangle}T$.
 For steady state it needs an inversion of
population $\lambda_{1}>\lambda_{0}$ and the condition $(1-B/\tan
B)>0$ due from the requirement of the stable equilibrium.

Then using (\ref{43}) under small fluctuations of intensity, one
finds the Mandel parameter
\begin{equation}\label{52}
\xi=\frac{T}{N} \frac{Q_{II}}{(2\lambda_{1}-1)\sin^{2}B(1-B/\tan B)}
\end{equation}
where the diffusion coefficient $Q_{II}$ has the form
\begin{eqnarray}\label{47}
Q_{II}&&\\
\nonumber
 =\frac{N}{T} \{-\frac{1}{2} \sin^{4} B +(2\lambda_{1}-1)B \sin
B\cos B &&\\
\nonumber
+(1-\lambda_{1})\sin^{2} B && \\
\nonumber + 2\lambda_{1}(1-\lambda_{1})\sin^{4} B &&\\
\nonumber
 +2(N-1)\lambda_{1}(1-\lambda_{1}) \sin^{4} B \}&&
\end{eqnarray}
In (\ref{47}) the atomic correlations of both state (\ref{1}) and
(\ref{2}) presented by the same term
$N2(N-1)\lambda_{1}(1-\lambda_{1})\sin^{4} b/T$ are non negative.
It means increasing of the light noise due from the initial atomic
correlations. This term is equal to zero, if $N=1$, or, if the
pump creates independent atoms. Indeed, the phase behavior is
insensitive to these atomic correlations. In accordance with the
small fluctuation approximation it needs $Q_{II}\ll \Gamma \langle
I \rangle^{2}$, that is the reason of restriction the number of
atom $N$.

Consider the noise of light, denoted as the spectrum of
photocurrent $i^{(2)}(\omega)$. It can be measured by a simple
heterodyne scheme included a detector and spectrum analyzer. At
frequencies  $\omega\approx 0$ the spectrum of the light noise
takes the form
\begin{equation}\label{53}
i^{(2)}(0)=1+2\xi\frac{1}{1-B/\tan B}
\end{equation}
where unit indicates the shot noise level or the standard quantum
limit, which is noise of the coherent field.
 The main features of the light statistics can be found from equation
(\ref{52}),(\ref{53}). If all atoms are prepared independently in
the state $|1\rangle^{\otimes N}$, then $\xi\geq -1$. The negative
Mandel parameter means, that in the cavity the photon statistics
is sub Poisson, and the noise of light can be suppressed below the
standard quantum limit. It $\xi=-1$, the photon number variance is
equal to zero, and one finds the Fock state of the light into
cavity $|n\rangle$, where $n=\langle I\rangle$. However the noise
reduction is small. Noise can be suppressed up to zero, when
$\xi=-1/2$. These results are in agreement with results for a
regular pumping, proposed in Ref. \cite{Golubev}. For the case of
correlated atoms, we discuss here, the Mandel parameter becomes
positive and the light noise increases.

In contrast, when a pump prepares atoms in the state given by
(\ref{3}), the atoms have the initial coherence $\langle
s_{01}(0)\rangle=
(\alpha\beta^{*}\delta_{b1}+\alpha^{*}\beta\delta_{b0})/\sqrt{2}=|\alpha\beta|\exp[i\varphi_{0}]$,
where $\varphi_{0}=(-1)^{(1-b)}\arg (\alpha\beta^{*})$ and
inversion $\langle s_{11}(0)\rangle-\langle
s_{00}(0)\rangle=(-1)^{1-b}|\alpha|^{2}$. Here $b=0,1$, that
indicates two version of the state (\ref{3}). If $b=0$ the
inversion of population is negative, in contrast the case when
$b=1$. Then together with the small fluctuation of the intensity,
assume the phase fluctuation is small in a sense, that  $\mu
=\varphi-\psi_{0}\ll\psi_{0}$. Then semiclassical equations for
the phase $\varphi_{0}$ and the intensity $\langle I\rangle$ take
the form
\begin{eqnarray}\label{54}
\frac{\partial}{\partial
t}\psi_{0}=-\sqrt{2}\frac{g}{\sqrt{I}}|\alpha\beta|\sin(\psi_{0}-\varphi_{0})&&\\
\label{541}
 \frac{\partial}{\partial t}\langle I\rangle
 =-C\langle I\rangle+\frac{2}{T}(-1)^{1-b}|\alpha|^{2}
 \sin^{2}B&&\\
\nonumber
 +\frac{2\sqrt{2}}{T}|\alpha\beta|\cos (\psi_{0}-\varphi_{0})\sin B \cos B &&
\end{eqnarray}
In the steady state $\psi_{0}=\varphi_{0}$ and this solution is stable. The
initial atomic coherence is presented by the last term in the right hand side
of (\ref{541}). It result in the steady state solution for intensity, even the
initial inversion of population, produced by the pump, is negative, when $b=0$.

Noise of the light or spectrum of the photocurrent at the
frequencies near zero can be written as
\begin{equation}\label{55}
i^{2}(0)=1+2\xi\frac{C}{\Gamma}
\end{equation}
where
\begin{eqnarray*}\label{56}
\Gamma=C(1-B \frac{(A ctg B+\sqrt{2}|\alpha\beta|(ctg^{2}B-1))}{(
A+\sqrt{2}|\alpha\beta| ctg B)})
\end{eqnarray*}
where $A=(-1)^{(1-b)}2|\alpha|^{2}$, $\xi=Q_{II}/(\langle I\rangle
\Gamma)$. The expression for the diffusion coefficients reads, if
$b=1$
\begin{eqnarray}\label{661}
Q_{II}&=&\frac{1}{T}[ +\frac{1}{2}A
(\sin^{4}B+2B \sin B \cos B)\\
\nonumber &+&\sqrt{2}|\alpha\beta|\cos 2B (B-\sin B \cos B)-Q]
\end{eqnarray}
if  $b=0$
\begin{eqnarray}\label{671}
\nonumber
Q_{II}&=&\frac{1}{T}[\frac{1}{2}A\sin^{2} B(\sin^{2} B+2)-2B\cos B\\
\nonumber
&+&\sqrt{2}|\alpha\beta|\cos 2B (B+\sin B \cos B)\\
&-&2\sin B \cos B-Q]
\end{eqnarray}
and
\begin{eqnarray}
Q&=&(2(1+2|\alpha|^{2})|\beta|^{2}\sin^{2} B \cos^{2} B\\
\nonumber &+&\sqrt{2}|\alpha\beta|A\sin B\cos B(1-\cos 2B)\\
\nonumber &-&\frac{1}{2}A^{2} \sin^{4} B
\end{eqnarray}
There are two types of steady states of the micromaser operation,
that can be pointed. For the first of them $b=0$, it means, that
atoms are prepared initially without inversion because $\langle
s_{11}(0)\rangle-\langle s_{00}(0)\rangle=-|\alpha|^{2}$. If
$|\alpha\beta|\approx 0.49$, then one finds light, which phase is
well defined and statistics of photons is sub Poisson. Indeed, the
light noise can be suppressed up to $i^{2}(0)\approx 0.27$, that
is limit here. The second type of operation is obtained, when
$b=1$. In this case the pump creates inversion of population of
the atomic levels and coherence. Atoms generate light, which level
of noise is various. Particularly, one finds the perfect noise
reduction, when $i^{2}(0)\approx 0$, if $|\alpha|^{2}\gg
|\beta|^{2}$. Physically, the last feature is clear, because the
initial state of atoms is close to a state, of two independent
atoms in the upper level. Indeed, in the micromaser operation the
ideal sub Poisson light can be achieved using a regular pumping of
atoms \cite{Golubev}. However note that to do it, atoms have to be
pumped by the light, which statistics of photons is already sub
Poisson \cite{G}.

This work was supported in part by the  Delzell Foundation Inc.
and RBRF grant committee.


\begin{thebibliography}{99}

\bibitem{GolSok}
Yu.M. Golubev, I.V. Sokolov. JETP {\bf 60}, 234 (1984).
\bibitem{Walls}
F. Haake, S.M. Tan, D.F. Walls.Phys. Rev. {\bf A34}, 4025, (1986).
\bibitem{Golubev}
Yu.M. Golubev. JETP {\bf 79}, 561 (1994).
\bibitem{MVadeikoRubinT}
G.P. Miroshnichenko, IP Vadeiko, AV Rubin, J. Timonen JETP Letters
{\bf 72}, 449 (2000)
\bibitem{LSc}
M.O. Scully, W.E. Lamb. Phys. Rev.{\bf 159}, 208 (1967).
\bibitem{FilLugSc}
P. Filipovicz, J. Javanainem. P Meystre. Phys. Rev.{\bf A34}, 4547
1986, J.Kraus, M.O. Scully, H. Walter. Phys.Rev {\bf A34} 2032
1986, L. Lugiato, M.O. Scully, H. Walter. Phys.Rev {\bf A36}, 740,
1987
\bibitem{BuzekHill}
V. Buzek, M. Hillery. Phys. Rev. {\bf A54}, 1844, (1996).
\bibitem{BuzBraunHill}
V. Buzek, S.L. Braunstein, M. Hillery, D. Drub. Phys. Rev. {\bf
A56}, 3446, (1997).
\bibitem{Gisin}
N. Gisin, S. Massar. Phys. Rev. Lett. {\bf 79}, 2153 (1997).
\bibitem{Vedral}
M. Murao, D. Jonathan, M.B. Plenio, V. Vedral. Phys. Rev.{\bf
A59}, 156 (1999).
\bibitem{Braunst}
S.L. Braunstein, P. van Loock. Multiuser quantum channels for
continuous variables. E-print, LANL, quant/ph 0012063 (2000).
\bibitem{Ourprep}
V.N. Gorbachev, A.I. Trubilko, A.A. Rodichkina, A.I. Zhiliba.
Journal QIC, 2, 367, (2002),  E-print, LANL, quant/ph 0011124
(2000).
\bibitem{Zurek}
W.K. Wotters, W.H. Zurek. Nature {\bf 299}, 802 (1982).
\bibitem{Lmcoh}
N. Lu, R. Bergou. Phys. Rev.{\bf A40}, 237 (1989).\\V.N. Gorbachev, A.I.
Trubilko. JETP {\bf 88}, 882 (1999).\\ V.N. Gorbachev, A.I. Trubilko. Optics
and Spectroscopy {\bf 89}, 766 (2000).
\bibitem{GZ} V.N. Gorbachev, A.I. Zhiliba. Quant. Opt.{\bf 5},
193 (1993).
\bibitem{G}
Yu.M. Golubev. JETP {\bf 107}, 401 (1995).
\end{thebibliography}
\end{document}